\documentclass[aps,prl,reprint,superscriptaddress]{revtex4-1}

\usepackage{color}
\usepackage{soul}
\usepackage{dcolumn}
\usepackage{graphicx}
\usepackage{amsmath}
\usepackage{amssymb}
\usepackage{bm}
\usepackage[normalem]{ulem}
\usepackage{subfiles}

\begin{document}
\mbox{}

\title{Observation of the Schmid-Bulgadaev dissipative quantum phase transition}

\author{R. Kuzmin}
\affiliation{Department of Physics, University of Wisconsin-Madison, Madison, WI 53706, USA.}
\email[]{rkuzmin@wisc.edu}
\affiliation{Department of Physics, University of Maryland, College Park, MD 20742, USA.}
\author{N. Mehta}
\affiliation{Department of Physics, University of Maryland, College Park, MD 20742, USA.}
\author{N. Grabon}
\affiliation{Department of Physics, University of Maryland, College Park, MD 20742, USA.}
\author{R. A. Mencia}
\affiliation{Department of Physics, University of Maryland, College Park, MD 20742, USA.}
\affiliation{École Polytechnique Fédérale de Lausanne, 1015 Lausanne, Switzerland.}
\author{A. Burshtein}
\affiliation{Raymond and Beverly Sackler School of Physics and Astronomy, Tel Aviv University, Tel Aviv 6997801, Israel.}
\author{M. Goldstein}
\affiliation{Raymond and Beverly Sackler School of Physics and Astronomy, Tel Aviv University, Tel Aviv 6997801, Israel.}
\author{V. E. Manucharyan}
\affiliation{Department of Physics, University of Maryland, College Park, MD 20742, USA.}
\affiliation{École Polytechnique Fédérale de Lausanne, 1015 Lausanne, Switzerland.}

\date{\today}
\begin{abstract}
\noindent Although quantum mechanics applies to many macroscopic superconducting devices, one basic prediction remained controversial for decades. Namely, a Josephson junction connected to a resistor must undergo a dissipation-induced quantum phase transition from superconductor to insulator once the resistor’s value exceeds $h/4e^2 \approx 6.5~\textrm{k}\Omega$ ($h$ is Planck's constant, $e$ is the electron charge). Here we finally demonstrate this transition by observing the resistor’s internal dynamics.  Implementing our resistor as a long transmission line section, we find that a junction scatters electromagnetic excitations in the line as either inductance (superconductor) or capacitance (insulator), depending solely on the line’s wave impedance. At the phase boundary, the junction itself acts as ideal resistance: in addition to elastic scattering, incident photons can spontaneously down-convert with a frequency-independent probability, which provides a novel marker of quantum-critical behavior.

\end{abstract}

\maketitle

Classical and quantum particles qualitatively differ in the way they propagate in a periodic potential: While a low-energy classical particle localizes inside a single potential minimum, a quantum one can avoid localization by tunneling to neighboring minima and forming extended states, as it happens to electrons in crystals~\cite{girvin2019modern}. Forty years ago, Schmid found that the two antagonistic behaviors are separated by a dissipation-induced quantum phase transition in the presence of a viscous friction~\cite{Schmid1983}. Such a transition has profound implications for superconducting Josephson junction devices, whose dynamics can also be modeled as a particle in a periodic potential (Fig.~1A,B). The macroscopic superconducting phase-difference variable $\varphi$ defines the particle's position, the potential is $-E_\textrm{J}\cos\varphi$, the particle’s mass is the junction’s self-capacitance $C_\textrm{J}$, and viscous friction can be provided by connecting a resistance $Z$. Localization in the $\varphi$-space corresponds to a classical Josephson effect, in which case the junction carries a supercurrent proportional to $\sin\varphi$. In an extended state, the supercurrent quantum-mechanically averages to zero, and the junction becomes an insulator. Bulgadaev further showed that the transition from superconductor to insulator is controlled neither by the Josephson energy $E_\textrm{J}$ nor by the charging energy $E_\textrm{C} = e^2/2C_\textrm{J}$, but only by the resistance $Z$, with the critical value given by a fundamental constant $R_Q = h/4e^2 \approx 6.5~\textrm{k}\Omega$~\cite{Bulgadaev1984}. Against common intuition, whether a superconducting weak link can carry a supercurrent is decided solely by its dissipative environment.

\begin{figure*}[ht]
	\centering
	\includegraphics[width=0.9\linewidth]{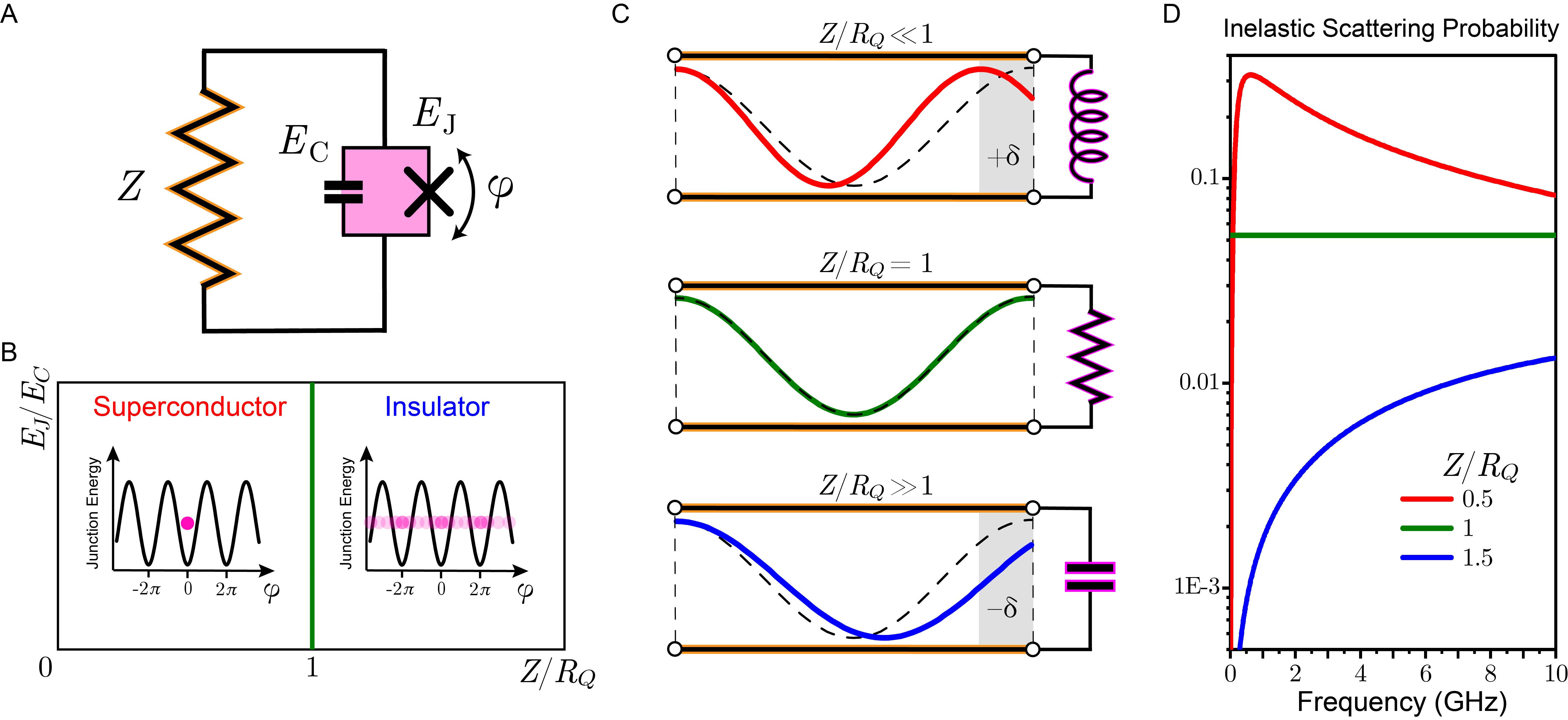}
	
	\caption{\textbf{Schmid-Bulgadaev transition viewed from ``inside" the resistor} \textbf{(A)} Resistively-shunted Josephson junction circuit. \textbf{(B)} Predicted phase-diagram, where superconductor (insulator) corresponds to localization (delocalization) of the junction's phase-difference $\varphi$ in the periodic Josephson potential. \textbf{(C)} Ohmic environment implemented as a long transmission line section with wave impedance $Z$. The same junction acts as inductive ($Z\ll R_Q$, superconductor), capacitive ($Z \gg R_Q$, insulator), or resistive ($Z = R_Q$, transition) termination of the line.
    \textbf{(D)} Probability for an incident environmental photon to scatter inelastically (down-convert to lower-frequency photons), calculated using an exactly-solvable model for $E_\textrm{J}/h=3~\textrm{GHz}$, $E_\textrm{C}/h=40~\textrm{GHz}$, and zero temperature (supplementary materials).
                }
		\label{fig:Fig1}
\end{figure*}

\begin{figure}[ht]
	\centering
	\includegraphics[width=1\linewidth]{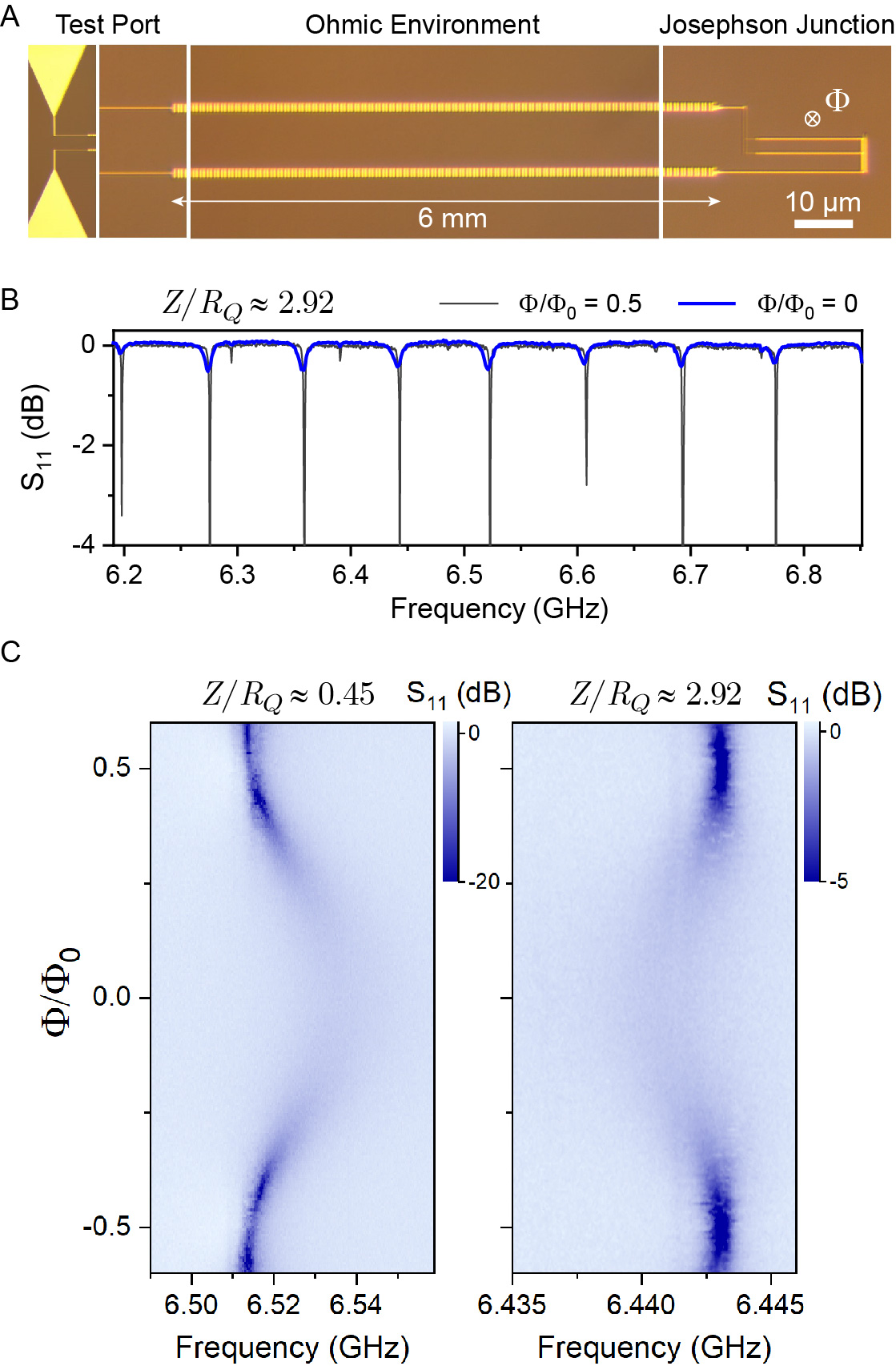}
	\caption{\textbf{Experiment implementation}. \textbf{(A)} Example device image: high-impedance transmission line terminated by a dipole antenna (left) and a split-junction (right), pierced by a flux $\Phi$. \textbf{(B)} Example of reflection spectroscopy data showing environmental modes unaffected ($\Phi/\Phi_0=0.5$) and maximally shifted ($\Phi/\Phi_0=0$) by the junction. \textbf{(C)} The magnitude of the reflection signal $S_{11}$ around one of the environmental modes as a function of the flux $\Phi$ for two devices with $Z<R_Q$ (left panel) and $Z>R_Q$ (right panel).
                } 
	
	\label{fig:Fig2}
\end{figure}

\begin{figure}[hb]
	\centering
	\includegraphics[width=0.9\linewidth]{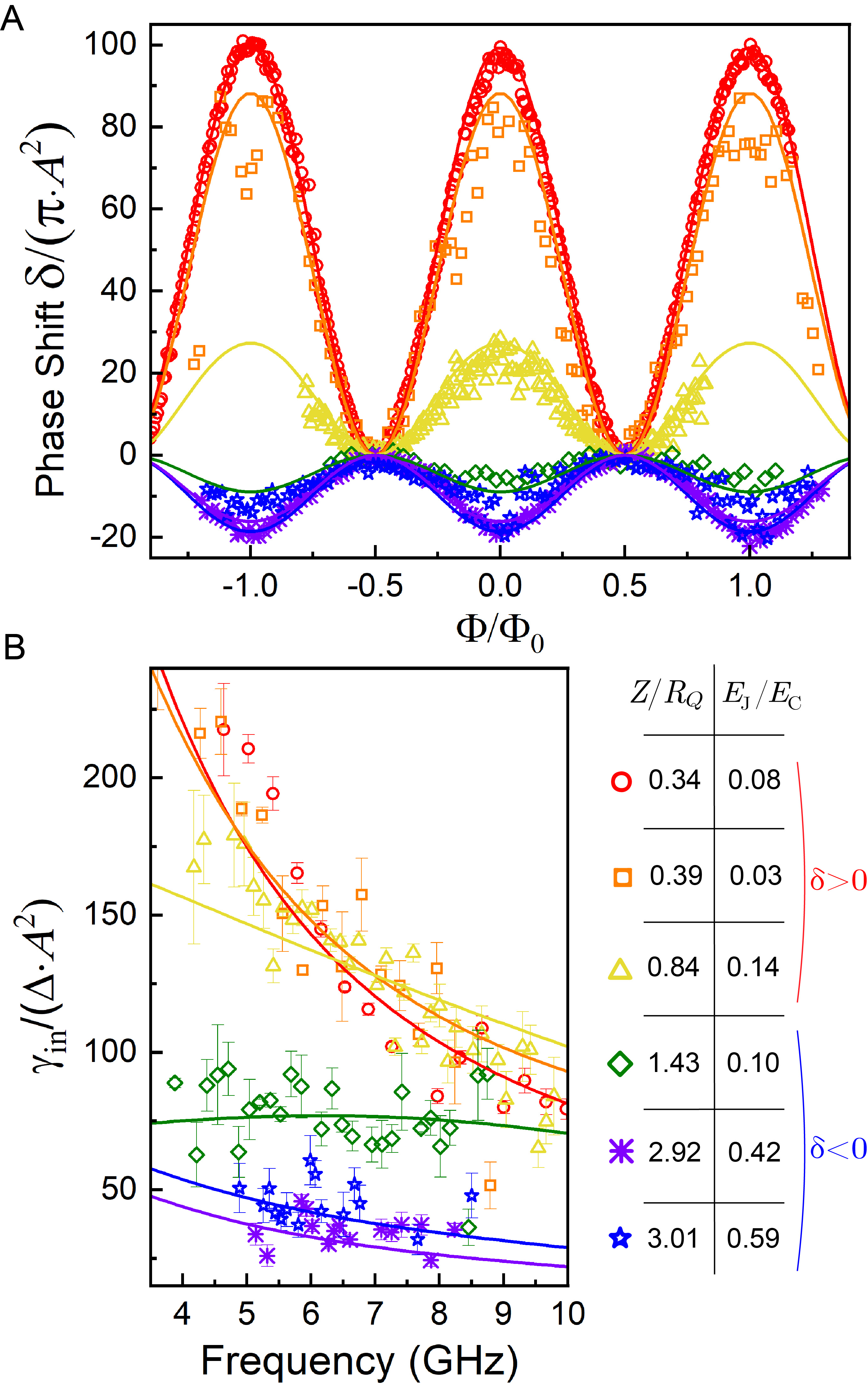}
	\caption{\textbf{Multi-device theory-experiment comparison}. \textbf{(A)} The experimental (markers) and theoretical (lines) results for the phase shift $\delta$ induced on environmental modes at $~6.7~\textrm{GHz}$ across multiple devices with varying impedances. \textbf{(B)} The experimental (markers) and theoretical (lines) results for the frequency dependence of the inelastic scattering rate across the same devices. Both the elastic and inelastic effects are normalized by the squared junction's area $A^2$. For details, see the supplementary materials.
                }
	
	\label{fig:Fig3}
\end{figure}

Although the theory behind Schmid-Bulgadaev (SB) transition was checked in multiple analytical~\cite{Aslangul1985,Guinea1985,Schon1990,Herrero2002,Kimura2004} and numerical~\cite{Werner2005,Lukyanov2007} works, it resisted experimental observation for decades. At the superconducting side of the phase-diagram, $Z \ll R_Q$, a dissipative stabilization of supercurrents in ultra-small Josephson junctions was well-understood classically~\cite{steinbach2001direct}. At the opposite side, junctions biased via high-impedance leads did show evidence of insulating DC transport~\cite{Cleland1990,Kuzmin1991Coulomb,Kuzmin1991BO,Watanabe2001}.
However, attempts to establish the transition via transport measurements proved inconclusive, at the very least suggesting that the critical resistance was not device-independent~\cite{Yagi1997,Penttila1999,Penttila2001,Watanabe2003}. A recent surge in revisiting this problem stirred even more controversy. An experiment measuring AC-impedance of resistively shunted junctions claims no transition at all~\cite{Murani2020}, sparking an ongoing debate~\cite{Hakonen2021, Murani2021, devoret2021does}. A bolometric heat transfer measurement also concludes the junction remains superconducting well past the SB-critical point~\cite{subero2022bolometric}. A new theory suggests the critical resistance must indeed deviate from $h/4e^2$ due to an overlooked effect of the charging energy contribution~\cite{Masuki2022}, causing yet another unresolved debate~\cite{Sepulcre2022,Masuki2022reply}. 
The disturbing \textit{status quo} is such that a standard quantum theory accounts for many complex superconducting qubit devices but apparently fails to describe a basic circuit in Fig.~1A, implying that we still lack a complete understanding of macroscopic quantum effects in superconductors.

\begin{figure*}[htpb]
	\centering
	\includegraphics[width=0.8\linewidth]{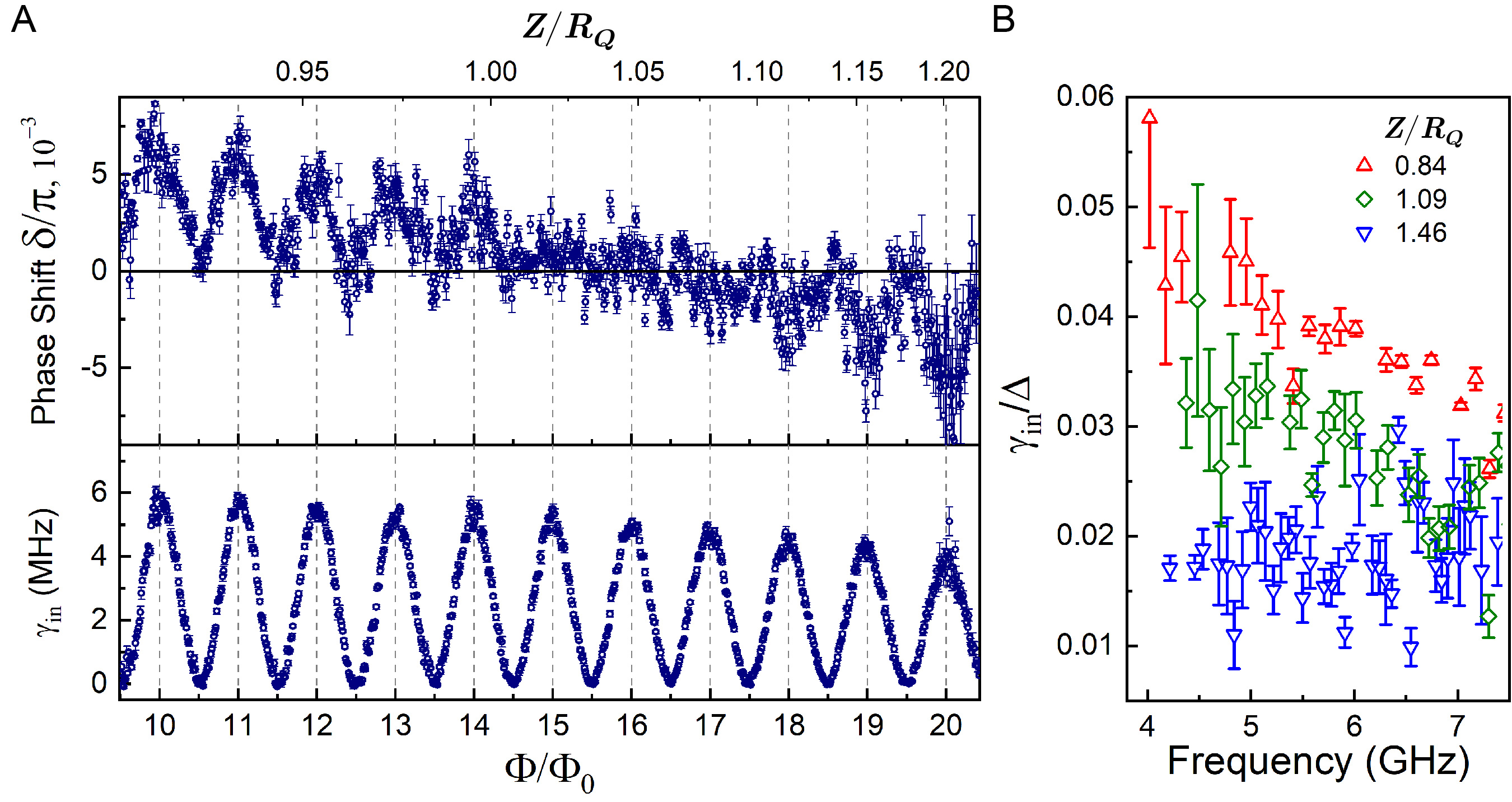}
	\caption{\textbf{Crossing the transition in a single device.} \textbf{(A)} Phase-shift (top) and inelastic scattering rate (bottom) measured at a frequency around $7~\textrm{GHz}$ as a function of global external flux, which in this device simultaneously modulates the junction's Josephson energy $E_\textrm{J}$ with a period $\Phi_0$ and tunes the wave impedance $Z$ across the transition. \textbf{(B)} Frequency dependence of the inelastic scattering rate at integer $\Phi/\Phi_0$ normalized by the free spectral range $\Delta$ at three values of $Z/R_Q$. 
                }
		\label{fig:Fig4}
\end{figure*}

Here we finally demonstrate the SB-transition by swapping the usual roles of the quantum system and its dissipative environment. Instead of asking how dissipation influences the Josephson phase-difference variable, we rather ask how the junction's dynamics acts back at the microscopic degrees of freedom of its environment. To this end, we recall that according to Caldeira and Leggett, a resistor $Z$ can be modeled as a long section of a telegraph-like transmission line with characteristic wave impedance $Z$~\cite{Caldeira1983}. Connecting a junction to such a resistor introduces an electrical termination in the line, which would scatter the environmental photons. With such a viewpoint from ``inside" the resistor, we consider the following signatures of the SB-transition:

 (i) In the zero-frequency limit, the elastic scattering phase-shift $\delta$ abruptly changes from $\delta =\pi/2$ at the superconducting side (short-circuit termination) to $\delta = 0$ at the insulating side (open-circuit termination).

 (ii) At a non-zero frequency, the short-(open-) circuit termination acts as inductance (capacitance), and hence the discontinuity in $\delta$ becomes a crossover from $\delta >0$ to $\delta <0$ (Fig. \ref{fig:Fig1}C). The capacitance originates from the formation of Bloch bands by the extended $\varphi$-variable, recently demonstrated in experiments on quasicharge qubits~\cite{Pechenezhskiy2020}. 

 (iii) Environmental photons can also scatter inelastically, the rate of which scales with frequency $f$ as $f^{2(Z/R_Q-1)}$ (Fig. \ref{fig:Fig1}D). This analytical result is valid for $f > E_\textrm{J}^*/h$, where $E_\textrm{J}^*$ is the Josephson energy renormalized by quantum fluctuations (the analog of Kondo temperature in similar models~\cite{gogolin2004bosonization}), satisfying $E_\textrm{J}^* \rightarrow 0$  as $Z \rightarrow R_Q^-$ and $E_\textrm{J}^* = 0$ for $Z \geq R_Q$ (supplementary materials). Note that inelastic scattering here does not involve any absorption. It is, in fact, a down-conversion of incident photons into lower-frequency photons due to the junction's non-linearity. While this effect is negligible deep enough into either the superconducting or insulating phase, it cannot be neglected close to the transition. At the critical point, the down-conversion probability becomes frequency-independent, i.e., each environmental mode sees the junction as effectively an ideal resistor approximately given by $R_Q(E_\textrm{C}/E_\textrm{J})^2$ for $E_\textrm{J} \ll E_\textrm{C}$ (Fig.~1C).

In experiments, the abrupt phase change (i) would be inevitably smeared by a finite temperature. However, the properties (ii, iii) are robust to a sufficiently low temperature $T$ as long as the probe frequency satisfies $f \gtrsim k_B T/h$ (supplementary materials). Likewise, the transition is insensitive to a finite size of the transmission line, as long as the free spectral range $\Delta$ of the standing-wave modes satisfies $\Delta \lesssim k_B T/h$. Finally, experiments have to be restricted to junctions with $E_\textrm{J}/E_\textrm{C} \lesssim 1$. Otherwise, the $\varphi$-tunneling becomes suppressed exponentially in $(E_J/E_C)^{1/2}$, and scattering data in the $\textrm{GHz}$-range is dominated by the Josephson plasma resonance~\cite{Kuzmin2021}. The transition would show up only at unrealistically low temperatures and long time scales~\cite{Houzet2020, Burshtein2021}.

We measured multiple devices in a broad range of wave-impedance $Z$ and the ratio $E_\textrm{J}/E_\textrm{C}$. Each device includes a section of an electromagnetic transmission line formed by two parallel arrays of 10,000 Josephson junctions (Fig. \ref{fig:Fig2}A) with sufficiently large areas to suppress their individual quantum phase-fluctuations~\cite{Watanabe2001arrays, Planat2019}. Such a line can readily reach wave impedance exceeding $R_Q$ while creating a bath of high-quality standing-wave modes \cite{Kuzmin2019SIT}. Shunting the line's end is a small Josephson junction in the form of a symmetric superconducting quantum interference device (SQUID). Its Josephson energy $E_\textrm{J}$ can be tuned by piercing the loop with an external magnetic flux $\Phi$. When the flux $\Phi$, normalized by the flux quantum $\Phi_0=h/2e$, takes half-integer values, the junction's $E_\textrm{J}$ is almost zero, effectively eliminating the junction. This method provides us with the baseline standing-wave modes in the waveguide corresponding to the zero scattering phase shift $\delta$ (see the black dashed line in Fig.\ref{fig:Fig1}C and data in Fig.~2B, black). Following a previously established procedure \cite{Kuzmin2019SIT}, we reliably extracted the impedance $Z$ and free spectral range $\Delta$ of all our transmission lines (see Table 1 in the supplementary materials).

Increasing $E_\textrm{J}$ by tuning the flux $\Phi$ towards an integer value, we observe the effect of the junction on the environmental modes, which acquire frequency shifts and broadening (Fig. \ref{fig:Fig2}B, blue).
Figure \ref{fig:Fig2}C zooms in on one of the modes (right) and compares its behavior with a mode in a device at the opposite side of the transition (left). The frequency shifts are in one-to-one correspondence with the phase shift $\delta$ at the junction's boundary, defined in units of $\pi$ as the frequency shift normalized by $\Delta$. Thus, an inductive boundary shifts modes up in frequency, whereas a capacitive boundary shifts them down (see Fig. \ref{fig:Fig1}C). It is then evident that the junction indeed behaves like an inductor with environmental impedance $Z/R_Q\approx0.45$ (positive shift) and like a capacitor in the opposite case $Z/R_Q\approx2.92$ (negative shift).

We summarize our data from different devices in Fig.~3 and compare it with calculations perturbative in the junction's Josephson energy $E_\textrm{J}$ (supplementary materials).
We define the inelastic scattering rate $\gamma_\textrm{in}$ as the difference between the mode's linewidth at integer and half-integer flux values, as seen in Fig. \ref{fig:Fig2}C.
To the lowest order in perturbation theory, the phase shift and the inelastic rate are quadratic in $E_\textrm{J}$. Because our experiment does not provide an easy way to measure $E_\textrm{J}$, we relied on its linear scaling with the junction's area $A$ and normalized both the phase shifts and the inelastic rates by $A^2$, thus removing the $E_\textrm{J}^2$ dependence. As a result, the data overlap for devices with similar impedance (Fig. \ref{fig:Fig3}), justifying such an approach and validating our perturbative treatment. The results of our calculations match the observed elastic (Fig. \ref{fig:Fig3}A) and inelastic (Fig. \ref{fig:Fig3}B) effects, using the impedance and the junction's Josephson and charging energies as adjustable parameters. The best-fit numbers agree with the measured $Z$ values and the junction dimensions (Table 1 in the supplementary materials). For example, the obtained scaling factors $124\pm15~\textrm{GHz}/\mu\textrm{m}^2$, which link $E_\textrm{J}/h$ to the junction's areas across all the devices, are close to the measured values in corresponding Josephson transmission lines and to junction arrays fabricated under similar conditions \cite{Kuzmin2019SIT}. A small deviation from the simple scaling behavior $f^{2(Z/R_Q-1)}$ for the two highest-impedance devices was explained by a combined effect of finite temperature and a lower high-frequency cut-off $\hbar/ZC_\textrm{J}$ (supplementary materials).

In a control experiment, we probed the transition in a single device, employing the tunability of Josephson junction arrays in a perpendicular magnetic field \cite{Kuzmin2022tuning}. With the same field, we controlled the junction's $E_\textrm{J}$ and tuned the environmental impedance across the $Z/R_Q=1$ point, measuring the phase shift and the inelastic rate at one of the modes (Fig. \ref{fig:Fig4}A top and bottom, respectively) (see supplementary materials for details). As in the experiment on multiple devices, the phase shift changes from positive to negative as the environment becomes high-impedance. The data shows that the transition happens in a narrow range of environmental impedances despite the fact the $E_\textrm{J}/E_\textrm{C}$ ratio is strongly modulated. Considering the measurement uncertainty on $Z$, we found the quantum-critical point at $Z/R_Q=1.0\pm0.1$.
Similarly, the effect of the environment on the inelastic rate $\gamma_\textrm{in}$ in a single device (Fig. \ref{fig:Fig4}B) agrees with the behavior observed across multiple devices (Fig. \ref{fig:Fig3}B). The trend of the frequency dependence of $\gamma_\textrm{in}$ changes direction when the impedance crosses the critical value $Z/R_Q=1$.

We conclude that a Josephson junction exposed to a properly engineered Ohmic dissipation does undergo a transition from a superconductor to an insulator, as originally predicted by Schmid and Bulgadaev. The critical environmental impedance $Z$ agrees with $R_Q = h/4e^2$ within our $10\%$ measurement uncertainty, and no influence of the junction parameters on the transition boundary was found over a factor of 20 variation range of $E_\textrm{J}/E_\textrm{C}$. For example, the weakest junction with $E_\textrm{J}/E_\textrm{C} = 0.03$ was superconducting at $Z/R_Q = 0.39$, while the strongest junction with $E_\textrm{J}/E_\textrm{C} = 0.59$ was insulating at $Z/R_Q = 3.01$. Note, even insulating response is generally flux-periodic, because nothing prevents Cooper pairs from crossing the junction back and forth with a zero net supercurrent, further highlighting the transition's macroscopic nature.

Our approach differs from previous attempts to observe a dissipative QPT in that we probe the quantum system (junction) using the same degrees of freedom (photons) that make up its environment. The environmental photons see a superconducting/insulating junction as an inductive/capacitive element away from the phase boundary, in agreement with common intuition. At the critical point, though, both the inductance and the capacitance must diverge due to growing quantum fluctuations, seemingly creating a paradox: at any finite frequency, the junction's apparent impedance would be simultaneously zero and infinite. This is where inelastic scattering processes come to the rescue. They make the junction appear as neither diverging inductance nor diverging capacitance, but an ideal (frequency-independent) resistance, thereby reconciling the two antagonistic phases. 

-----------------------------------

\textbf{Acknowledgements.} We acknowledge the support from DOE Early Career Award (DE-SC0020160) and US-Israel Binational Science Foundation through Grant No.~2020072.
A.B. and M.G. were further supported by the Israel Science Foundation (ISF) and the Directorate for Defense Research and Development (DDR\&D) Grant No.~3427/21.

\bibliography{DQPT}

\end{document}